\documentclass[12pt]{article}
\usepackage{amsmath,amssymb,epsfig}
\usepackage{graphicx,floatflt}
\usepackage{epstopdf}
\usepackage{cite}

\def\cF{\mathcal{F}}
\def\bF{\bar{F}}
\def\bW{\bar{W}}
\def\bpsi{\bar{\psi}}
\def\blm{\bar{\lambda}}
\def\bcF{\bar{\mathcal{F}}}

\def \cN{\mathcal{N}}
\def \cL{\mathcal{L}}

\def \be{\begin{equation}}
\def \ee{\end{equation}}
\def \beq{\begin{eqnarray}}
\def \eeq{\end{eqnarray}}

\makeatletter

\renewcommand\section{\@startsection {section}{1}{\z@}%
                                   {-3.5ex \@plus -1ex \@minus -.2ex}%
                                   {2.3ex \@plus.2ex}%
                                   {\normalfont\large\bfseries}}

\renewcommand\subsection{\@startsection{subsection}{2}{\z@}%
                                     {-3.25ex\@plus -1ex \@minus -.2ex}%
                                     {1.5ex \@plus .2ex}%
                                     {\normalfont\normalsize\bfseries}}

\newcount\hour \newcount\minute
\hour=\time \divide \hour by 60
\minute=\time
\def\now{%
\ifnum \hour<13
  \ifnum \hour=0 \advance \hour by 12 \number\hour:\else \number\hour:\fi%
     \ifnum \minute<10 0\fi%
     \number\minute%
\ A.M.%
\else \advance \hour by -12 \number\hour:%
  \ifnum \minute<10 0\fi%
  \number\minute%
  \ P.M.%
\fi%
}

\makeatother

\begin{document}

% format
\baselineskip=18pt  % a la harvmac
\numberwithin{equation}{section}  % make eq labels (sec.num)
\allowdisplaybreaks  % allow page breaks in displayed eqs

% print date, time and filename
%\pagestyle{myheadings}
%\markright{{\tt \jobname.tex} -- \today{} \now}

%%%%%%%%%%%%%%%%%%%%%%%%%%%%%%%%%%%%%%%%%%%
%%%        TITLE BEGINS HERE
%%%%%%%%%%%%%%%%%%%%%%%%%%%%%%%%%%%%%%%%%%%

%% ========== title (note version) begins here ==========
%
%\vspace*{-1cm}
%\begin{center}
% {\Large\bf Title of the Document}
%\end{center}
%\vspace*{-.5cm}
%
%% ========== title (note version) ends here ==========

%% ========== title (paper version, a la harvmac) begins here ==========

\thispagestyle{empty}

% Report number
\vspace*{-2cm}
\begin{flushright}
{\tt arXiv:0806.4733}\\
CALT-68-2692\\
IPMU-08-0038
\end{flushright}

% title, authors, affiliation
\vspace*{1.7cm}
\begin{center}
 {\large {\bf Current Correlators for General Gauge Mediation}}

 \vspace*{2cm}
 Hirosi Ooguri,$^{*,\dagger}$
 Yutaka Ookouchi,$^*$ Chang-Soon Park,$^*$ and Jaewon Song$^*$\\
 \vspace*{1.0cm}
 $^*$
{\it California Institute of Technology 452-48, Pasadena, CA 91125, USA}\\
~~\\
 $^\dagger$ {\it Institute for the Physics and Mathematics of the Universe, \\
University of Tokyo, Kashiwa, Chiba 277-8582, Japan}\\[1ex]
 \vspace*{0.8cm}
% {\tt foo@bar},\begin{thebibliography}{1}
% {\tt sh{}i{}ge{}@t{}h{{}}{e{ory}}.c{}a{}l{{}}te{{}}ch.e{{}}d{}u} % can this avoid spam?
\end{center}
\vspace*{.5cm}

% abstract
\noindent
In the gauge mediation mechanism, the effects of the hidden sector are characterized by a set of correlation functions of the global symmetry current of the hidden sector. We present methods to compute these correlators in cases with strongly coupled hidden sectors. Several examples are presented to demonstrate
the technique explicitly.

\newpage
\setcounter{page}{1} % don't number title page

%% ========== title (paper version, a la harvmac) ends here ==========

%%%%%%%%%%%%%%%%%%%%%%%%%%%%%%%%%%%%%%%%%%%\lim_{x->0}
%%%           TITLE ENDS HERE
%%%%%%%%%%%%%%%%%%%%%%%%%%%%%%%%%%%%%%%%%%%

%\tableofcontents
%\printindex

%%%%%%%%%%%%%%%%%%%%%%%%%%%%%%%%%%%%%%%%%%%
%%%        MAIN TEXT BEGINS HERE
%%%%%%%%%%%%%%%%%%%%%%%%%%%%%%%%%%%%%%%%%%%

\section{Introduction}
Gauge mediation has been an attractive mechanism for mediating the supersymmetry breaking effects.
%It transfers the SUSY breaking effects via fields charged under the Standard Mo%del gauge group. Compared to other methods of mediation, it has a number of goo%d features such as flavor universality, predictability. But they also have some% problems such as the $\mu$ problem. However it wasn't clear whether these feat%ures are generic to the gauge mediation or they are specific to a given model. %Also, usual gauge mediation mechanisms have a predictive power only when a weak%ly-coupled description is available.
Recently, Meade, Seiberg and Shih \cite{Meade:2008wd} presented a very general definition of the gauge mediation mechanism that includes most models with the supersymmetry breaking sector and the messengers, and models with the direct mediation. They also computed several physical quantities such as gaugino and sfermion masses on a general ground. Many of the physically observable effects mediated to the MSSM can be encoded by the current-current correlators of the global symmetry of the hidden sector. This allows us to extract the soft supersymmetry breaking terms even in the strongly coupled hidden sector\footnote{For an earlier work for gauge mediation with strongly coupled hidden sector, see \cite{NTWY}. }.

In this paper, we provide methods to compute the zero-momentum current correlators in some classes of strongly coupled gauge theories. The way we calculate the current correlators is to weakly couple the system to some `spectator' gauge theory. Sometimes the coupled system is solvable, in which case, by taking the decoupling limit, we are able to obtain useful information about the current correlators.

We will review the idea of general gauge mediation of \cite{Meade:2008wd} briefly and then understand that the correlators of the global currents essentially characterize a wide range of gauge mediation models. Then we go on to the basic scheme of our computation of current correlators. To illustrate the technique, we provide a number of computable examples.

%%%%%%%%%%%%%%%%%%%%%%%%%%%%%%%%%%%%%%%%%%%%%%%%%%%%%%%%%
\section{Review of General Gauge Mediation}

In this section, we briefly review the definition of the gauge mediation and the determination of the effect of the hidden sector via current correlations in \cite{Meade:2008wd}. Readers who are familiar with the subject may skip this section safely. In \cite{Meade:2008wd}, a careful definition of the gauge mediation is given. Following their definition, a model has the gauge mediation mechanism if the theory decouples into the MSSM and a separate hidden sector that breaks SUSY in the limit the MSSM gauge couplings $\alpha_i$ all vanish. In this setup, we may be able to compute various quantities in perturbation theory in the gauge coupling $\alpha_i$, but the hidden sector may be strongly coupled. The information of the hidden sector, however, can be parameterized by the current correlation functions in the hidden sector. In the supersymmetric gauge theory, a global current superfield $\mathcal{J^A}$ has the component form
\begin{equation}
\mathcal{J^A}=J^A+i \theta j^A -i\bar{\theta}\bar{j}^A-\theta \sigma^{\mu} \bar{\theta} j_{\mu}^A + \frac 1 2 \theta^2 \bar{\theta} \bar{\sigma}^{\mu} \partial_{\mu} j^A - \frac 1 2 \bar{\theta}^2 \theta \sigma^{\mu} \partial_{\mu} \bar{j}^A - \frac 1 4 \theta^2 \bar{\theta}^2 \square J^A\;,
\end{equation}
and satisfies the current conservation conditions
\begin{equation}
\bar{D}^2 \mathcal{J}^A=D^2 \mathcal{J}^A=0
\end{equation}
with $j_{\mu}$ satisfying $\partial^{\mu} j^A_{\mu}=0$. Here $A$ is an index for the adjoint representation of the global symmetry group. We follow the notation of \cite{Wess:1992cp}. The current-current correlators have the following general forms:
\begin{equation}
\begin{split}
\langle J^A(x) J^B(0) \rangle &= \delta^{AB} \frac 1 {x^4} C_0 (x^2 M^2)\\
\langle j^A_{\alpha}(x) \bar{j}^B_{\dot{\alpha}}(0)\rangle&= -i \delta^{AB}  \sigma^{\mu}_{\alpha \dot{\alpha}} \partial_{\mu} \left({1\over  x^4} C_{1/2} (x^2 M^2)\right)\\
\langle j^A_{\mu}(x) j^B_{\nu}(0)\rangle &= \delta^{AB} (\eta_{\mu\nu} \partial^2 - \partial_{\mu} \partial_{\nu})\left(\frac 1 {x^4} C_1(x^2 M^2)\right)\\
\langle j^A_{\alpha}(x) j^B_{\beta}(0) \rangle &= \delta^{AB} \epsilon_{\alpha\beta} \frac 1 {x^5} B_{1/2} (x^2 M^2)\;,
\end{split}
\end{equation}
where $M$ is the characteristic mass scale of the theory. $B_{1/2}$ may be complex but all $C_a$ are real. There could also be nonzero one-point function $\langle J(x) \rangle$, but it vanishes for nonabelian currents. When supersymmetry is not broken spontaneously, we have the relations
\begin{equation}
C_0=C_{1/2}=C_1\;,\qquad \text{and} \qquad B_{1/2}=0\;.
\end{equation}
Since supersymmetry is restored in UV, whether SUSY is spontaneously broken or not,
\begin{equation} \label{BCUV}
\lim_{x \rightarrow 0} C_0(x^2 M^2)=\lim_{x \rightarrow 0} C_{1/2}(x^2 M^2) =\lim_{x \rightarrow 0} C_1(x^2 M^2)\;,\qquad \text{and} \qquad \lim_{x \rightarrow 0} B_{1/2}(x^2 M^2)=0\;.
\end{equation}
We now gauge the global current by coupling it to a gauge field. The part of the original Lagrangian for the gauge field is given by
\begin{equation}
\begin{split}
\mathcal{L}&=\frac{1}{8\pi} \mathrm{Im} \left(\tau \mathrm{Tr} \int d^2 \theta^2 W^{\alpha} W_{\alpha} \right)+\cdots\\
	   &=\frac 1 {2g^2} D^A D^A-\frac i {g^2} \lambda^A \sigma^{\mu} D_{\mu} \bar{\lambda}^A-\frac 1 {4g^2} F^A_{\mu\nu} F^{A\mu\nu}+\cdots\;,
\end{split}
\end{equation}
where $\mathrm{Im} \tau= \frac{4\pi}{g^2}$ and we use normalization such that $\mathrm{Tr} T^A T^B=\delta^{AB}$.
After integrating out the hidden sector, its effect is determined by the current correlation functions. Here we will only consider one gauge group and not a product of gauge groups, such as in the MSSM, for simplicity. Ignoring higher derivative terms, the change of the effective Lagrangian is
\begin{equation}\label{E:ChangeL}
\delta \mathcal{L}_{eff}=\frac 1 2 \tilde{C}_0 (0) D^A D^A - \tilde{C}_{1/2}(0) i \lambda^A \sigma^{\mu} \partial_{\mu} \bar{\lambda}^A - \frac 1 4 \tilde{C}_1(0) F^A_{\mu\nu} F^{A\mu\nu}-\frac 1 2(M \tilde{B}_{1/2}(0) \lambda^A \lambda^A + c.c.)+\cdots\;.
\end{equation}
Here $\tilde{C}_a$ and $\tilde B$ are Fourier transforms of $C_a$ and $B$, respectively:
\begin{equation}
\begin{split}
\tilde{C}_a\left(\frac{p^2}{M^2};\frac{M}{\Lambda}\right)&=\int d^4 x e^{ipx} \frac 1 {x^4} C_a(x^2 M^2)\\
M\tilde{B}_{1/2}\left(\frac{p^2}{M^2}\right)&=\int d^4 x e^{ipx} \frac 1 {x^5} B_{1/2}(x^2 M^2)\;.
\end{split}
\end{equation}
The gaugino and sfermion masses are also determined by the current correlation functions. The gaugino mass at tree level can be read off from the change of the Lagrangian \eqref{E:ChangeL}:
\begin{equation}\label{E:Gaugino}
M_{\lambda} = g^2 M \tilde{B}_{1/2} (0)\;.
\end{equation}
When the one point function $\langle J \rangle$ is zero, the sfermion mass occurs at one loop. In this case, we have to know the current correlation functions at the momentum of order $M$, the typical scale of the hidden sector. The sfermion mass is then
\begin{equation}
m^2_{\tilde{f}}=g^4 c_{2f} A\;,
\end{equation}
where $c_{2f}$ is the quadratic Casimir of the representation of $f$ under the gauge group and
\begin{equation}
A=-\int \frac{d^4 p}{(2\pi)^4} \frac 1 {p^2} \left(3\tilde{C}_1 \left(\frac{p^2}{M^2}\right)-4\tilde{C}_{1/2}\left(\frac{p^2}{M^2}\right)+\tilde{C}_0\left(\frac{p^2}{M^2}\right)\right)\;
\end{equation}

%%%%%%%%%%%%%%%%%%%%%%%%%%%%%%%%%%%%%%%%%%%%%%%%%%%%%%%

\section{Basic Idea}
In this section, we present the basic idea to compute the current correlators in non-supersymmetric vacua, which encode the gaugino and sfermion masses. Schematically, the full theory can be written in the following form:
\be
 \cL = \cL_{hid} + \cL_{int} + \cL_{MSSM}\;,
\ee
where $\cL_{hid}$ is the supersymmetry breaking hidden sector, $\cL_{MSSM}$ is the visible MSSM sector, and $\cL_{int}$ is the interaction between the two sectors, which transmits the supersymmetry breaking effects to the MSSM. When we integrate out $\cL_{hid} + \cL_{int}$, we produce the soft terms that break supersymmetry in the MSSM Lagrangian.
\be
 \cL \to \cL_{eff} = \cL_{MSSM} + \delta \cL_{soft}\;.
\ee
This can be done explicitly if the hidden sector is weakly coupled, but it is hard to do so for the strongly coupled case in general.

However, we can circumvent the difficulties in certain cases. As we have reviewed in the previous section, many of the parameters of the soft terms can be determined by calculating the current correlation functions. To achieve this, we will replace the MSSM with another theory such that we have control over the combined theory. That is, we are going to consider the following Lagrangian:
\be
 \cL_{g'} = \cL_{hid} + \cL_{int}' + \cL_{spec}'\;.
\ee
Here, if the gauge coupling constant $g'$ for the spectator Lagrangian goes to zero, the spectator fields decouple. Supposing this new theory is solvable, we integrate out the hidden sector and obtain
\be
 \cL_{g', eff} =  \cL_{spec}' + \delta \cL_{soft}\;.
\ee
Now, by taking $g^{\prime}$ small and extracting nontrivial terms, we get the desired soft terms for the original theory.

In the following sections, we will be more specific and gauge the flavor symmetry of the hidden sector. Suppose the hidden sector is a gauge theory with gauge group $G_1$ and global symmetry group $G_2$, and each field lives in the representation $(A_i, B_i)$ under $G_1 \times G_2$. Now by weakly gauging $G_2$, we get the theory with the product gauge group $G_1 \times G_2$:
\be \label{LagHid}
 \cL = \cL_{G_1} + \cL_{int} + \cL_{G_2}\;.
\ee
Let the dynamical scales associated with gauge groups $G_1$ and $G_2$ be
$\Lambda_1$ and $\Lambda_2$, respectively, and $g'$ gauge coupling for the
gauge group $G_2$. We set the scale $\Lambda_1 \gg \Lambda_2$, so that we can treat the gauge interaction for $G_2$ to be very weak at the scale we probe. If we can integrate out the whole Lagrangian \eqref{LagHid} in this limit, we are able to extract information about the current correlation functions of the flavor symmetry of the hidden sector. Note that this probe Lagrangian should not be confused with the Lagrangian for the real visible sector, although the structure is very similar. Here, this is just a probe to extract the information of the hidden sector. In the next section, we provide certain classes of examples in which we can follow this procedure.

Before we go on, let us briefly mention that we do not really have to gauge the full global symmetry. It is actually enough to gauge any subgroup $H$ of the global symmetry group $G$, because the global currents are only sensitive to group theory factors. To see this, suppose the global current for $G$ is in a representation $R$. Let the generators of $G$ be $\{ T^a \}$ and $H$ $\{ t^\alpha \}$. The global current correlators can be written as
\be
 \langle J^a (x) J^b (0) \rangle = f(x) \delta^{ab}\;,
\ee
where we have omitted Lorentz indices. This is the only possible form by symmetry. The current correlators come with the group theory factor
\be
 \langle J^a (x) J^b (0) \rangle_G = \tilde{f}(x) D_G(R) \delta^{ab}\;,
\ee
where $D_G(R)$ is the Dynkin index of the group $G$ in the representation $R$ and we have omitted the Lorentz indices. Now, let's decompose the representation in terms of the representations of the subgroup $H$ by $R = \oplus_i r_i $ where $r_i$ is a representation of $H$. Then, the correlators we get by gauging the subgroup can be written as
\be
 \langle J^\alpha (x) J^\beta (0) \rangle_H = \tilde{f}(x) \sum_i D_H(r_i) \delta^{\alpha \beta}.
\ee
Therefore, we can multiply the currents obtained by gauging the subgroup by the group theory factors to get the desired current correlators.
\be
 \langle J^a (x) J^b (0) \rangle_G = \frac{D_G(R)}{\sum_i D_H(r_i)} \langle J^a (x) J^b (0) \rangle_H.
\ee

%%%%%%%%%%%%%%%%%%%%%%%%%%%%%%%%%%%%%%%%%%%%%%%%%%%%%%

\section{Examples}

\subsection{$\cN=2$ SQCD with a superpotential}

Here we will consider $\mathcal{N}=2$ $SU(N_c)$ Seiberg-Witten theory with $N_f$ hypermultiplets\cite{Seiberg:1994rs, Seiberg:1994aj} with an appropriate superpotential as the hidden sector. This theory has $U(N_f)$ global symmetry. We will weakly gauge the $SU(N_f)$ part of the global symmetry. So the hidden sector is given by the Lagrangian
\begin{equation}\label{E:SWmicro}
\begin{split}
\mathcal{L}=&\frac{1}{8\pi}\mathrm{Im} \left[ \tau \left(\mathrm{Tr} \int d^2 \theta W^{\alpha} W_{\alpha} +2 \int d^2 \theta d^2\bar{\theta} \Phi^{\dagger} e^{-2V} \Phi\right)\right] + \int d^2 \theta W(\Phi) + c.c. \\
&+\int d^2\theta d^2\bar{\theta} \left(Q_a^{\dagger} e^{-2V} Q_a + \tilde{Q_a} e^{2V} \tilde{Q_a}^{\dagger} \right)+\int d^2\theta \left(\sqrt{2} \tilde{Q_a} \Phi Q_a + M \tilde{Q_a} Q_a \right) + c.c\;,
\end{split}
\end{equation}
where $a=1,\cdots,N_f$ and $W(\Phi)$ is a superpotential for the adjoint chiral superfield $\Phi$. Note that $M$ in the mass term $M\tilde{Q_a} Q_a$ is proportional to the $N_f \times N_f$ identity matrix, which is of the most general form to preserve $SU(N_f)$ global symmetry.
We will consider the Coulomb branch of the hidden sector, whose special coordinates are denoted by $a^i$ where $i=1, \cdots, N_c$ with $\sum a^i =0$. With a suitable choice of the superpotential $W(\Phi)$, the hidden sector can be in a metastable SUSY breaking state.
For example, we can use a K\"ahler normal coordinate truncated at some finite order as a superpotential to get a metastable SUSY breaking state at a generic point in the Coulomb branch \cite{Ooguri:2007iu,Marsano:2007mt}. Note that $a^i$ may have nonzero $F$-component, which we denote by $F^i$. Our purpose is to compute the current-current correlators of the $SU(N_f)$ global current below the typical scale of $a^i$.

To achieve this, we gauge the flavor symmetry of the hypermultiplets. Since our purpose is to calculate the current-current correlators, we need not gauge the flavor symmetry using the MSSM. To facilitate computations, it is better to gauge the flavor symmetry by another $\mathcal{N}=2$ $SU(N_f)$ SW gauge theory. So the total Lagrangian can be written as
\begin{equation}\label{E:TotalL}
\begin{split}
\mathcal{L}'=&\frac{1}{8\pi}\mathrm{Im} \left[ \tau \left(\mathrm{Tr} \int d^2 \theta W^{\alpha} W_{\alpha} +2 \int d^2 \theta d^2\bar{\theta} \Phi^{\dagger} e^{-2V} \Phi\right)\right] + \int d^2 \theta W(\Phi) + c.c. \\
&+\int d^2\theta d^2\bar{\theta} \left(Q^{\dagger} e^{-2V-2V'} Q + \tilde{Q} e^{2V+2V'} \tilde{Q}^{\dagger} \right)\\
&+\int d^2\theta \left(\sqrt{2} \tilde{Q_a} \Phi Q_a + M \tilde{Q_a} Q_a +\sqrt{2} \tilde{Q_a} \Phi'_{ab} Q_b\right) + c.c.\\
&+\frac{1}{8\pi}\mathrm{Im} \left[ \tau' \left(\mathrm{Tr} \int d^2 \theta W'^{\alpha} W'_{\alpha} +2 \int d^2 \theta d^2\bar{\theta} \Phi'^{\dagger} e^{-2V'} \Phi'\right)\right]\;,
\end{split}
\end{equation}
where primes denote similar multiplets in the $SU(N_f)$ gauge theory and the trace in the last line is over the flavor $SU(N_f)$ indices. Also, in the second line, the first term may be explicitly expressed as
\begin{equation}
Q^{\dagger i}_a \left(e^{-2V}\right)_i^j \left(e^{-2V'}\right)^a_b Q_j^b\;,
\end{equation}
where $i,j=1,\cdots,N_c$ and $a,b=1,\cdots, N_f$.
The gauge coupling of the spectator gauge theory is assumed to be very weak, so $\tau'=4\pi i /g'^2$ with $g'$ very small.

Treating the theory as $\mathcal{N}=2$ $SU(N_c)\times SU(N_f)$ SW gauge theory, the low energy effective theory in the Coulomb branch is given by
\begin{equation}\label{E:ProdGage}
\begin{split}
\mathcal{L}'_{eff}=&\frac{1}{8\pi}\mathrm{Im} \left[\int d^2 \theta \left(\mathcal{F}_{ij} W^{\alpha i} W^j_{\alpha} + 2 \mathcal{F}_{ia} W^{\alpha i} W'^{a}_{\alpha} + \mathcal{F}_{ab} W'^{\alpha a} W'^{b}_{\alpha}\right) \right]\\
&+\frac{1}{4\pi}\mathrm{Im} \left[\int d^2 \theta d^2\bar{\theta} \left(\mathcal{F}_i \bar{a}^i + \mathcal{F}_{a} \bar{m}^{a}\right)\right]\;,
\end{split}
\end{equation}
where $a^i$ and $m^{a}$ are eigenvalues of $\Phi$ and $\Phi'$, respectively, and subscripts under $\mathcal{F}$ denote differentiations. Note that the prepotential $\mathcal{F}$ depends both on $a^i$ and $m^{a}$. Usually, it is hard to compute $\mathcal{F}$ for product gauge groups. However, using the fact that the spectator gauge theory is weakly coupled, we may obtain necessary information out of $\mathcal{F}_0$ of the original $\mathcal{N}=2$ $SU(N_c)$ SW theory. To see this, note that the low energy effective theory of \eqref{E:SWmicro} is
\begin{equation}\label{E:OrigLow}
\mathcal{L}_{eff}=\frac{1}{8\pi}\mathrm{Im} \left[\int d^2 \theta \mathcal{F}_{0ij} W^{i\alpha} W^j_{\alpha} +2 \int d^2 \theta d^2\bar{\theta}\mathcal{F}_{0i} \bar{a}^i \right]\;,
\end{equation}
where $\mathcal{F}_0(a)$ is the prepotential for $\mathcal{N}=2$ $SU(N_c)$ SW theory. Let us consider the case where $g^{\prime}$ goes to zero. In such a limit, $m^{a}$ is treated as constant. Note that the dynamics of $a^i$ in \eqref{E:ProdGage} and \eqref{E:OrigLow} agree when $\mathcal{F}_{i}=\mathcal{F}_{0i}$ and $m^{a}\sim 0$. $m^{a}\sim 0$ is necessary since we set all nonabelian mass parameters to vanish in \eqref{E:OrigLow}. The condition $m^{a}\sim 0$ really means that the typical scale of $m^{a}$ is much smaller than that of $a^i$. That is, of the moduli space of the Coulomb branch of the product gauge group, in the region where $|a| \gg |m|$, we may interchange $\mathcal{F}_i$ and $\mathcal{F}_{0i}$ freely. Therefore, if our interest is to consider the low energy effective theory whose typical scale of $m$ is much lower than that of $a$, we may use the prepotential of the original $\mathcal{N}=2$ $SU(N_c)$ SW theory to compute quantities.

In the Seiberg-Witten curve language, one can obtain the prepotential in this limit by taking appropriate mass deformation of the curve, and regarding it as the moduli of the gauge theory. To see this, let's compare \eqref{E:SWmicro} and \eqref{E:TotalL}. In the limit where the second gauge group $SU(N_f)$ decouples, the only difference is the term $\sqrt{2} \tilde{Q_a} \Phi'_{ab} Q_b$ in the superpotential. The adjoint scalar component of $\Phi'_{ab}$ acts as a mass term for the hypermultiplets. Therefore, by identifying the massive deformation parameter $\tilde{m^a}$ to be $\sqrt{2}m^a$, and computing the period integrals, we can obtain the prepotential $\cF$ and its derivatives in the limit of $|m| \ll |a|$.

\subsubsection{Calculation of $\tilde{B}_{1/2}(0)$}

From \eqref{E:ProdGage}, we can read off the coefficients of $D'^2$, $\lambda' \sigma^{\mu} \partial_{\mu} \bar{\lambda'}$, $F'_{\mu\nu} F'^{\mu\nu}$ and $\lambda' \lambda'$ of the spectator gauge theory. Let's first consider the coefficient of $\lambda' \lambda'$. To show what is going on explicitly, let us write down the Lagrangian in terms of the component fields ignoring terms involving derivatives:
\begin{equation}\label{E:Leff}
\begin{split}
\mathcal{L}'_{eff}=\cdots+&g_{IJ} F^I \bF^J+\frac 1 {2\pi} g_{IJ} D^I D^J + \frac 1 {8\sqrt{2}\pi} \cF_{IJK} D^I \psi^J \lambda^K+\frac 1 {8\sqrt2\pi} \bcF_{IJK} D^I \bpsi^j \blm^K \\
&+\frac{i}{16\pi} \cF_{IJK} \bF^I \psi^J \psi^K - \frac{i}{16\pi} \cF_{IJK} F^I \lambda^J \lambda^K+c.c.\\
&+\frac{i}{32\pi} \cF_{IJKL}(\psi^I\psi^J)(\lambda^K\lambda^L)+W_I F^I-\frac 1 2 W_{IJ} \psi^I \psi^J+c.c.
\end{split}
\end{equation}
Here $I=(i,a)$ and the prepotential $\cF$ has the superfields $A^I=(a^i, m^{a})$ as its arguments. Note that $W_{a}=0$. $g_{IJ}=\frac 1 {4\pi} \mathrm{Im} \cF_{IJ}$ is the metric for the product group gauge theory.
After integrating out the auxiliary fields $F^I$, we can read off the coefficients of $\lambda^I \lambda^J$ and these are given by
\begin{equation}\label{E:LambdaI}
\mathcal{L}'_{eff}=\cdots+\frac i {16\pi} \cF_{IJK} g^{KL} \bW_L \lambda^I \lambda^J+\cdots\;.
\end{equation}
If the hidden sector is very weakly interacting with the spectator gauge theory, $g^{KL} \bW_L$ should nearly be the same as $(g_0^{ij} \bW_j, 0)$ where $g_0^{ij}$ is the inverse metric of the original $\mathcal{N}=2$ $SU(N_c)$ gauge theory. To see this, note that among the components of the metric $g_{IJ}$, $g_{ab}$ is very large compared with $g_{ij}$ and $g_{ia}$ in the limit $g'\rightarrow 0$. From the expression below
\begin{equation}
\delta^K_J=g_{IJ} g^{JK} = \begin{pmatrix} g_{ij} & g_{ib} \\ g_{aj} & g_{ab} \end{pmatrix} \begin{pmatrix} g^{jk} & g^{jc} \\ g^{bk} & g^{bc} \end{pmatrix}\;,
\end{equation}
we see that $g^{ij}$ is much larger than $g^{ib}$ and $g^{ab}$ and this is more so as $g'\rightarrow 0$. Also $g^{ij}$ is the inverse of $g_{ij}$, which is the same as $g_{0ij}$, all in the limit $g'\rightarrow 0$. Therefore, $g^{aL} \bW_L$ can be ignored compared to $g^{ij} \bW_j$ and
\begin{equation}\label{E:limF}
g^{IJ} \bW_J=(g_0^{ij} \bW_j, 0)=:(-F^i, 0)\;.
\end{equation}
Next, we need to integrate out $\lambda^i$ fields in \eqref{E:LambdaI}. When we do this, we get
\begin{equation}\label{E:Lambdac}
\mathcal{L}=\cdots-\frac i {16\pi} \left(F^i \cF_{iab} - (\cF_{ijm} F^m)^{-1} \cF_{aik} \cF_{bjl}F^{k}F^{l}\right) \lambda^{a}\lambda^{b}+\cdots\;,
\end{equation}
where $(\cF_{ijm} F^m)^{-1} $ is the inverse of the matrix $X_{ij}=(\cF_{ijm} F^m)$ and we use the relation \eqref{E:limF} in our limit. There are two terms in the coefficient of $\lambda^{a}\lambda^{b}$. The first term is the usual one, but the second is from integrating out the gaugino $\lambda^{i}$ in the hidden sector. However, when we gauge the $SU(N_f)$ flavor current, the second term can be neglected. The point is that, in the second term, $\cF$ is differentiated only once by $m^{a}$. Note that $\cF_0$ in the original $\mathcal{N}=2$ $SU(N_c)$ SW theory depends smoothly on the symmetric polynomials of $m^{a}$'s, such as $\sum_{a\le b} m^{a} m^{b}, \sum_{a\le b\le c} m^{a} m^{b} m^{c}, \cdots$. Hence, when differentiated once, $\cF_0$ necessarily contains at least one factor of $m^{a}$ and so goes to zero when $|m|/|a|\rightarrow 0$. On the other hand, $\cF_0$ itself or its second derivatives by $m^{a}$ need not vanish when $m^{a}$ is small. Since $\cF_{0i}=\cF_{i}$ in
 our limit, in the region of the moduli space where $|a|\gg |m|$, the first term dominates and we can safely ignore the second term.
Therefore, when the hidden sector fields are integrated out, we may say that the relevant gaugino mass term is
\begin{equation}\label{E:Lambdap}
\mathcal{L}=\cdots-\frac i {16\pi} F^k \cF_{kab} \lambda^{a} \lambda^{b}+\cdots\;.
\end{equation}
Note that, if we gauged the $U(N_f)$ flavor current, the prepotential would depend on $\sum_{a} m^{a}$ and we would not be able to ignore the second term in \eqref{E:Lambdac}.

Next, we will look for a theory at the scale of $|a|$ which gives the low energy effective theory described above. That is, we hide all signals of the hidden sector into the gauge coupling of the spectator gauge theory and consider $\mathcal{N}=2$ $SU(N_f)$ SW theory
\begin{equation}\label{E:HidIntOut}
\mathcal{L}=\frac{1}{8\pi}\mathrm{Im} \left[ \tau_2 \left(\mathrm{Tr} \int d^2 \theta W'^{\alpha} W'_{\alpha} +2 \int d^2 \theta d^2\bar{\theta} \Phi'^{\dagger} e^{-2V'} \Phi'\right)\right]\;,
\end{equation}
where $\tau_2=4\pi i /g'^2+\theta^2 F_{\tau}$ is the gauge coupling of the probe Lagrangian at the scale of $|a|$.

This is a theory at the scale of $|a|$. We will go to the low energy effective theory in the Coulomb branch where the superfield $\Phi'$ has eigenvalues $m^{a}$, $a=1,\cdots, N_f$(so $\sum m^{a}=0$). Although $|a|\gg |m|$, the coupling $g'$ is so small that one loop correction is enough when we consider the dynamics at the scale of $|m|$. The prepotential at one loop is given by
\begin{equation}
\cF(m)= \frac{\tau_2}{2} \sum_{a} \left(m^{a} - \frac{\sum_{b} m^{b}}{N_c}\right)^2 +\frac i {4\pi} \sum_{a<b}(m^{a}-m^{b})^2 \log\frac{(m^{a}-m^{b})^2}{\Lambda^2}\;,
\end{equation}
and the low energy effective theory is
\begin{equation} \label{E:HidOutLow}
\mathcal{L}=\frac{1}{8\pi}\mathrm{Im} \left[\int d^2 \theta \mathcal{F}_{ab} W'^{\alpha a} W'^{b}_{\alpha}+2 \int d^2 \theta d^2\bar{\theta} \mathcal{F}_{a} \bar{m}^{a}\right]\;.
\end{equation}
Note that the $\theta^2$ component of $\cF$ is not corrected by the one-loop effect. So the gaugino mass term is given by
\begin{equation}
\mathcal{L}=\cdots-\frac 1 {8\pi} F_{\tau} \lambda^{a} \lambda^{a}+\cdots\;.
\end{equation}
Note that we use coordinates for the second gauge group such that $\sum_{a=1}^{N_f} \lambda^{a}=0$.
Comparing this with \eqref{E:Lambdap}, we obtain
\begin{equation}
F_{\tau} \left(\delta_{ab}-\frac 1 {N_c}\right) = \frac i 2 F^k \mathcal{F}_{kab}
\end{equation}
at $m^{a}=0$. Note that, since $\cF_k$ depends on the masses $m^a$ by the combination $m^a - \sum_b m^b/Nc$, taking derivatives with respect to $m_a$ and $m_b$, $F_{kab}$ in the right-hand side has the same index structure as that in the left-hand side. From \eqref{E:ChangeL},
\begin{equation} \label{E:BExp}
M \tilde{B}_{1/2}(0)\left(\delta_{ab}-\frac 1 {N_c}\right) = \frac i {8\pi} F^k \cF_{0kab}\;.
\end{equation}

It is instructive to actually calculate the gaugino mass using this formula in the semiclassical regime.
That is, we check the expressions in the case where the expectation value of the chiral superfield $\Phi$ of the hidden sector is much larger than the scale of the hidden sector gauge theory. Additionally, we assume that the hypermultiplets $Q$ and $\tilde{Q}$ are massless: i.e. $M=0$ in \eqref{E:TotalL}. The chiral superfield $\Phi_{ij}$ has nonzero $F$-term $F_{\Phi ij}$ where $i$ and $j$ are the gauge group indices for the hidden sector. Note that we are in the Coulomb branch of the hidden sector. We use gauge transformation such that $\langle \Phi_{ij}\rangle$ is diagonal with diagonal elements $a_i$($\sum a_i =0$). Let $F_i$ be the corresponding $F$-term of $a_i$. We will calculate the gaugino mass in this setup. We start with the Lagrangian \eqref{E:TotalL} and go to the low energy effective theory at the scale of $|m|$, where $\langle \Phi' \rangle$ has eigenvalues of $m_{a}$ with the constraint $\sum m_{a}=0$. The $a_i$ dependent part of the prepotential $\mathcal{F}$
 , given by \cite{Ennes:1998ve} with the constraint $\sum m_{a}=0$ built in, is
\begin{equation}\label{E:SCPre}
\begin{split}
\mathcal{F}=&\frac {2\pi} {g^2} \sum_i \left( a_i - \frac {\sum_j a_j}{N_c}\right)^2\\
	    &+ \frac i {4\pi} \sum_{i<j} \left(a_i - a_j \right)^2 \log \frac {(a_i-a_j)^2}{\Lambda^2}\\
	    &- \frac i {8\pi} \sum_{i,b} \left(a_i - m_{b} - \frac {\sum a} {N_c}+\frac{\sum m} {N_f} \right)^2 \log \frac {\left(a_i - m_{b} - \frac {\sum a} {N_c}+\frac{\sum m} {N_f} \right)^2}{\Lambda''^2}+\cdots\;,
\end{split}
\end{equation}
where $\tau=4\pi i /g^2$, $\tau'=4\pi i/g'^2$. $\Lambda$ is the scale for the first gauge group and $\Lambda''$ is some scale which does not change the answer that follows. Then
\begin{equation}\label{E:DDDFL}
\mathcal{F}_{kab}=-\frac i {2\pi} \sum_i \frac 1 {a_i-m_{b}} \left(\delta_{ik}-\frac 1 {N_c}\right)\left(\delta_{ab}-\frac 1 N_f\right)\;,
\end{equation}
after imposing tracelessness conditions.
Since the scale of the spectator gauge theory does not enter into the expression, we may set $\mathcal{F}_{0kab}=\mathcal{F}_{kab}$. In the limit $m_{b}\rightarrow 0$, we have, from \eqref{E:BExp},
\begin{equation}
M\tilde{B}_{1/2}(0)=\frac{1}{16\pi^2} \sum_k \frac {F_k}{a_k}\;,
\end{equation}
which gives the usual one-loop gaugino mass through \eqref{E:Gaugino}
\begin{equation}
M_{\lambda}=\frac{g^2}{16\pi^2} \sum_k \frac {F_k}{a_k}\;,
\end{equation}
if we identify $\sqrt 2 a_k$ with masses of the messengers(The $F$-term of $\sqrt 2 a_k$ is $\sqrt 2 F_k$).

\subsubsection{Calculation of $\tilde{C}_a(0)$ and sfermion masses}
Using the same technique, we may compute $\tilde{C}_a(p^2/M^2)$ at zero momentum. Note that the effect of integrating out the hidden sector fields is to change the gauge coupling $\tau'$ to $\tau_2$ as shown in \eqref{E:HidIntOut}. Also, in this case, all $\tilde{C}_a(0)$ are the same. Using \eqref{E:ChangeL},
\begin{equation}\label{E:Ca}
\tilde{C}_a(0)=\frac 1 {4\pi} \mathrm{Im} \left(\tau_2(a)-\tau'\right)\;.
\end{equation}
When we go to the low energy effective theory, $\tau_2(a)$ gets renormalized. But since $g'$ is very small, it is enough to consider only the one-loop effect. So $\tau_2(a)$ gets only additive renormalization of the form $\log m$, which is independent of $a$. Hence we have
\begin{equation} \label{E:CExp}
\frac{\partial \tau_2(a)}{\partial a^k}\left(\delta_{ab}-\frac 1 N\right)=\cF_{kab}\;.
\end{equation}
Integrating this equation, we are able to obtain $\tau_2(a)$, which then may be fed into \eqref{E:Ca} to get $\tilde{C}_a(0)$. The additive constant is determined by noting that $\tilde{C}_a(0)$ goes to zero in the limit where $a\rightarrow \infty$.

Since we are required to calculate $\tilde{C}_a(p^2/M^2)$ when $p^2/M^2$ is of order 1, the information we have just obtained is not enough to calculate the sfermion masses of the MSSM. Alternatively, we can introduce a matter multiplet charged with respect to $SU(N_f)$ and evaluate its low energy effective action terms. A work in this direction is in progress.

\subsubsection{Generalization to other gauge groups}
Let us briefly comment on how the expressions \eqref{E:BExp} and \eqref{E:CExp} change for other non-abelian groups. We start from \eqref{E:HidIntOut}. Let $T_{A}$ be a basis of the adjoint representation of the flavor symmetry group and $H_{a}$ be a basis of the Cartan subalgebra. The prepotential as a function of the $\mathcal{N}=2$ adjoint chiral superfield $\Phi'$ at classical level is given by
\begin{equation}
\mathcal{F}(\Phi')=\frac 1 2 \frac{\tau_2}{c_{adj}} \mathrm{Tr}_{adj} (\Phi'^2)\;,
\end{equation}
where $c_{adj}$ is the Dynkin index for the adjoint representation:
\begin{equation}
\mathrm{Tr_{adj}}(T_{A} T_{B})=c_{adj}\delta_{AB}\;.
\end{equation}
In the Coulomb branch, all massive modes are integrated out and $\Phi'$ is diagonalizable due to $D$-term constraint. Hence $\Phi'=\sum_{a} m^{a} H_{a}$ and the prepotential above becomes
\begin{equation}
\mathcal{F}(a)=\frac 1 2 \frac{\tau_2}{c_{adj}} \mathrm{Tr}_{adj} (H_{a} H_{b})m^{a} m^{b}=\frac{\tau_2}{c_{adj}} m^{a} m^{b} \sum_{\alpha} \alpha_{a} \alpha_{b} \;,
\end{equation}
where the summation is over all positive roots of the flavor group.
Therefore $\lambda \lambda$ term in \eqref{E:HidOutLow} becomes
\begin{equation}
\mathcal{L}=\cdots-\frac 1 {4\pi} \frac {F_{\tau}} {c_{adj}} \sum_{\alpha} \alpha_{a} \alpha_{b} \lambda^{a} \lambda^{b}\cdots\;.
\end{equation}
We now compare this with the $\lambda\lambda$ term in the low energy theory of the product gauge group \eqref{E:Lambdap}. Therefore
\begin{equation}
F_{\tau}\sum_{\alpha} \alpha_{a}\alpha_{b} = \frac{i c_{adj}}{4} F^k \mathcal{F}_{kab}\;.
\end{equation}
Since the $\theta^2$ component of $\tau$ does not receive corrections at one loop, this $F_{\tau}$ can be used to calculate $M\tilde{B}_{1/2} (0)$. That is, the part of $\lambda\lambda$ consisting of the Cartan subalgebra part in \eqref{E:ChangeL} is
\begin{equation}
\delta \mathcal{L}_{eff}=\cdots-M \tilde{B}_{1/2}(0) \frac 1 {c_{adj}} \sum_{\alpha} \alpha_{a} \alpha_{b} \lambda^{a} \lambda^{b}\cdots\;.
\end{equation}
Therefore the relation corresponding to \eqref{E:BExp} is
\begin{equation} \label{NABExp}
M\tilde{B}_{1/2}(0) \sum_{\alpha} \alpha_{a} \alpha_{b} = \frac i {16\pi} c_{adj} F^k \mathcal{F}_{0kab}\;.
\end{equation}
Similarly, corresponding to \eqref{E:CExp}, we have
\begin{equation} \label{NACExp}
\frac{\partial \tau_a (a)}{\partial a^k} \sum_{\alpha} \alpha_{a} \alpha_{b} =\frac 1 {8\pi} c_{adj} \cF_{0kab}\;.
\end{equation}

\subsubsection{Hypermultiplet condensation}
It may be of interest to check whether the hypermultiplet bilinear $\tilde{Q}_a Q_b$ develops a nonzero expectation value. The Lagrangian for the product $SU(N_c)\times SU(N_f)$ gauge theory \eqref{E:TotalL} has some terms containing the $F$ component of $\Phi'$:
\begin{equation}
\mathcal{L}'=\cdots+\frac 1 {g'^2} \mathrm{Tr}\left( \bar{F}_{\Phi'} F_{\Phi'}\right)+\tilde{Q}_a F_{\Phi' ab} Q_b+c.c.+\cdots\;.
\end{equation}
Therefore we can calculate the expectation values of the bilinear $\tilde{Q}_a Q_b$ by differentiating the partition function with respect to $F_{\Phi'}$. More precisely, we will calculate the traceless part of $\langle \tilde{Q}_a Q_b \rangle$ since the source $F_{\Phi'}$ is traceless.
To get the effective Lagrangian for the spectator gauge theory, we start with \eqref{E:Leff} and set all vev's for the fermions to zero. Then we integrate out the $F$ and $D$ terms for the hidden sector. The equation of motion for $D^i$ sets $D^i=0$. The equation of motion for $F^i$ is
\begin{equation}
\bar{F}^{\bar j}=-g^{i\bar j}g_{i\bar a} \bar{F}^{\bar a}-g^{i\bar j} W_i\;,
\end{equation}
Plugging the result into the Lagrangian, we get
\begin{equation}
\mathcal{L}'_{eff}=-g^{j\bar i}(g_{a\bar i} F^{a}+\bar{W}_{\bar i})(g_{j\bar b}\bar{F}^{\bar b} + W_j)+g_{a\bar b} F^{a}{\bar F}^{\bar b}\;.
\end{equation}
Now we can read off the linear term in $F^{a}$:
\begin{equation}
\begin{split}
\mathcal{L}'_{eff}&=\cdots-g^{j\bar i}g_{a\bar i} W_j F^{a}+\cdots\\
		  &=\cdots-g_0^{j\bar i}g_{0a\bar i} W_j F^{a}+\cdots\;,
\end{split}
\end{equation}
where the second line follows in the limit $g'\rightarrow 0$. But as we argued below \eqref{E:Lambdap}, $g_{i \bar a}$ vanishes as $m^{a}$ goes to $0$ since we differentiate the prepotential $\mathcal{F}$ only once with respect to the second gauge indices to get $g_{i \bar a}$. Therefore, the linear term vanishes and the hypermultiplet bilinear can have at best an expectation value of the form
\begin{equation}\label{E:QQh}
\langle \tilde{Q}_a(0)Q_b(0)\rangle = h(a) \delta_{ab}\;,
\end{equation}
for some function $h(a)$. This does not break $U(N_f)$ symmetry. Also, $Q$ or $\tilde{Q}$ cannot have nonzero expectation values perturbatively in the superpotential $W$. The reason is that we have a $U(1)$ subgroup of $SU(2)$ $R$-symmetry of $\mathcal{N}=2$ theory even after including a superpotential if we assign charge +2 to the superpotential. Since $Q$ and $\tilde{Q}$ have charge +1, it cannot be expressed as a series in $W$. Therefore, although there could be hypermultiplet condensation, $U(N_f)$ symmetry is still preserved. Note that, if we gauged $U(N_f)$ symmetry instead, the bilinear would be diagonal with $a$-th diagonal element $g^{i\bar j} g_{i\bar a} W_j$. In this case, this would not vanish when $m^{a} \rightarrow 0$. However, by symmetry, $g_{i\bar a}$ would be the same for all $a$ for a fixed $i$. Hence the vev of the bilinear would be proportional to the identity matrix, and $U(N_f)$ symmetry would still exist. Of course, the answer does not depend on which
 global symmetry we gauge. Therefore $h(a)$ in \eqref{E:QQh} is determined and we have, for any index $c$,
\begin{equation}\label{E:QQ}
\langle \tilde{Q}_a(0)Q_b(0)\rangle = g_0^{j\bar i}g_{0c\bar i} W_j \delta_{ab}\;, \text{for the metric $g_0$ of $U(N_f)$}\;.
\end{equation}
Note that we actually calculate $\langle \tilde{Q}_a(0)Q_b(0)\rangle$ at the scale of $|m|$. But $\langle \tilde{Q}_a(0)Q_b(0)\rangle$ both at the scale of $|m|$ and at the scale of $|a|$ have the same form since when we go from the scale $|a|$ to $|m|$, we receive only perturbative effects, and this does not change $\langle \tilde{Q}_a(0)Q_b(0)\rangle$.

Having derived the formula for the quark condensate \eqref{E:QQ}, let us verify it in the semiclassical regime.
The leading contribution of $\sqrt2 \tilde{Q}^{i}_a F_{\Phi i}^j Q^{a}_{j}$ to $\langle\tilde{Q}^{i}_a Q^{b}_{i}\rangle$ for small $F_{\Phi}$, shown in Figure 1, is

%\begin{figure}
%\begin{center}
%\includegraphics[width=6cm]{fig1.pdf}
%\end{center}
%\end{figure}

\begin{figure}[t]
\begin{center}
  \epsfxsize=7.0cm \epsfbox{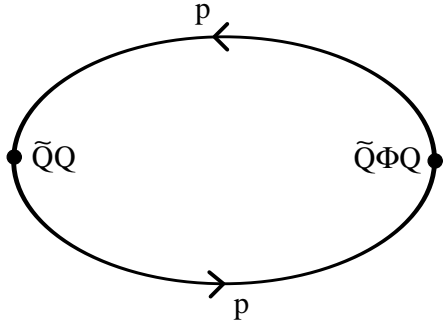}
\caption{\sl
 One loop diagram contributing hypermultiplet condensation}
\end{center}
\end{figure}

\begin{equation}\label{E:QQdiagram}
\begin{split}
\langle \tilde{Q}^i_{a} Q^b_i \rangle&=\sum_{i} \int_0^{\Lambda_0} \frac {d^4 p}{(2\pi)^4} \frac {\bar{F}_i \delta^b_a}{(p^2+(\sqrt 2 a_i)^2)^2}\\
				     &=-\frac 1 {16\pi^2} \delta^b_a \sum_{i} \bar{F}_i \log(\sqrt 2 a_i)^2\;.
\end{split}
\end{equation}
Note that the interaction is insensitive to the cutoff $\Lambda_0$. Let's compare this with \eqref{E:QQ}. The relevant part of the prepotential is similar to \eqref{E:SCPre}:
\begin{equation}
\begin{split}
\mathcal{F}=&\frac {2\pi} {g^2} \sum_i (a_i)^2\\
	    &+ \frac i {4\pi} \sum_{i<j} \left(a_i - a_j \right)^2 \log \frac {(a_i-a_j)^2}{\Lambda^2}- \frac i {8\pi} \sum_{i,a} \left(a_i - m_{a} \right)^2 \log \frac {(a_i-m_{a})^2}{\Lambda''^2}+\cdots\;,
\end{split}
\end{equation}
Note we do not impose the constraint $\sum m_{a}=0$ since \eqref{E:QQ} is valid when gauging $U(N_f)$ symmetry. The metric component $g_{ia}$ at weak coupling is given by
\begin{equation}
g_{a\bar i}=\frac 3 {16\pi^2} + \frac 1 {16\pi^2} \log \frac {(a_i -m_a)^2}{\Lambda''^2}\;.
\end{equation}
Let's go to the limit $m^{a}\rightarrow 0$. Then $g_0^{j\bar i} W_j = -\bar F^{\bar i}$ and using \eqref{E:QQ},
\begin{equation}
\langle \tilde{Q}^k_{a} Q^{b}_k \rangle = -\bar{F}^{\bar k} g_{0c \bar k} \delta^{b}_{a} = -\frac 1 {16\pi^2} \delta^{b}_{a} \sum_k \bar F_k \log(\sqrt 2 a_k )^2\;.
\end{equation}
The result agrees with \eqref{E:QQdiagram}.

%%%%%%%%%%%%%%%%%%%%%%%%%%%%%%%%%%%%%%%%%%%%%%%%%%%%%%%%%%%%%%%%
\subsection{Geometrically Realized Models}
%%%%%%%%%%%%%%%%%%%%%%%%%%%%%%%%%%%%%%%%%%%%%%%%%%%%%%%%%%%%%%

As another controllable model, we study geometrically induced supersymmetry breaking configuration in Type IIB string theory on $A_2$-fibered geometry. This has been studied in \cite{ABF} somewhat in a different context. Consider $A_2$ fibred geometry \cite{Cachazo} defined by
$${
x^2+y^2+ z \left(z-m_1 (t-a_1)\right)\left(z+m_2(t-a_2)\right)=0.
}$$
There are three singular points, $t=a_{1,2}$ and $a_3=(m_1a_1+m_2a_2)/(m_1+m_2)$. Wrapping $N_c$ anti-$D5$ and $N_f$ $D5$ branes on two ${\bf S}^2$s that resolve the singularities at $t=a_1$ and $a_2$ respectively, we can construct a supersymmetry breaking configuration. We do not wrap any brane at $t=a_3$, because this can decay into a lower energy configuration. Our present setup does not have an unstable mode as has been discussed in \cite{Mukhi}. Therefore we can take field theory limit. Here we claim that as long as the size of ${\bf S}^2$ at $t=a_2$ is much bigger than that at $t=a_1$, there is a field theory description for this brane/anti-brane system. According to the conjecture proposed in \cite{ASBVI,ASBVII,ABF,ABK}, there is a glueball description with respect to the hidden sector gauge group corresponding to a partial geometric transition. Thus it is reasonable to claim that the low energy field theory description is an interacting product $U(1)\times U(N_f)$ gau
 ge theory. The kinetic term is
\begin{eqnarray}
\int d^4 \theta K(S_1)\Phi_2 \Phi^{\dagger}_2 +{\rm Im} \left[ \int d^4 \theta \bar{S}_1{\partial {\cal F}_{0,0}(S_1)\over \partial S_1} \right] + \int d^2 \theta \left[ \tau \left(S_1 \right) W^{\alpha }W_{\alpha} \right]+c.c+\cdots
\end{eqnarray}
where $\cdots$ includes higher derivative terms and $U(1)$ gauge kinetic terms. ${\cal F}_{0,0}(S_1)$ is the prepotential for the geometry after the transition, \begin{equation}
2\pi i {\cal F}_{0,0} = {S_1^2\over 2} \left[ \log \left(S_1 \over m_1\Lambda_0^2 \right) -{3\over 2}\right]. \nonumber
\end{equation}
where $\Lambda_0$ is a cutoff scale of this description. The superpotential terms are
\begin{equation}
W=\alpha S_1 +N_c {\partial {\cal F}_{0,0} \over \partial S_1}+ \tilde{W}_2 (\Phi_2, S_1) \label{flavorSup}
\end{equation}
In application to phenomenology we will identify a subgroup of $SU(N_f)$ as the standard model gauge group. So the adjoint field $\Phi_2$ for $SU(N_f)$ gauge group should be integrated out by taking $m_2 \to \infty$. In the limit, the superpotential $\tilde{W}_2$ becomes a relatively simple function,
$${
\tilde{W}_2=\int_{\Lambda_0}^{a_2} \left[{m_1\over 2}(t-a_1)+{1\over 2}\sqrt{m_1^2 (t-a_1)^2-{4S_1 \over m_1}}  \right]\, dt .
}$$

Our goal in this section is to compute the function $\tau (S_1)$ and extract $C_i$s and $B_{1/2}$ from it. In the open string description we can say that this $S_1$ dependence is generated by the bifundamental matter. On the other hand, after the transition in closed string point of view, it is generated by closed string modes. To extract the interacting part, we use the glueball description for $U(N_f)$ gauge group as well
%$${
%S_2={1\over 32 \pi^2} \tr W^{\alpha}W_{\alpha } ={1\over 32 \pi^2 }\tr \lambda %\lambda+{1\over 32 \pi^2}\theta^2 \tr \left[-2i \lambda \sigma^{\mu} \partial_{%\mu} \bar{\lambda}-{1\over 2}F_{\mu\nu}F^{\mu \nu}+D^2 +{i\over 4}F^{\mu \nu} F%^{\rho \sigma}\epsilon_{\mu \nu \rho \sigma} \right].
%}$$
and assume that the glueball fields and $U(1)\subset U(N_f)$ gauge supermultiplet $w_{\alpha}$ are background fields. Following \cite{DVI,DVV}, we use glueball description for evaluating the interacting part even though the $SU(N_f)$ theory is weakly coupled and is not confined. Turning on these backgrounds modifies the geometry slightly. At the leading order of the modification, we read off the kinetic term for the gauge group. The low energy description is given by
\begin{eqnarray}
{\cal L}&=&{\rm Im} \left(\int d^4\theta  \bar{S}_i {\partial {\cal F}_0 \over \partial S_i} +\int d^2 \theta {1\over 2} {\partial {\cal F}_0 \over \partial S_i \partial S_j }w_i w_j \right)+\int d^2 \theta W(S_i)+c.c \label{def}
\end{eqnarray}
where $W(S_i)$ is Gukov-Vafa-Witten superpotential\cite{Gukov:1999ya} generated by the flux.
Solving the equation of motion for $F^1$, we obtain the potential
\begin{equation}
V={1\over g_{11}} \big| g_{12}\bar{F}^2+\partial_1 W \big|^2 -g_{22}F^2\bar{F}^2 -F^2 \partial_2 W -\bar{F}^2 \bar{\partial}_2 \bar{W}, \label{Poten}
\end{equation}
where we ignored $U(1)$ gauge fields. The metric is defined by ${\rm Im}\, \partial_i \partial_j {\cal F}_{0}$. Since we are interested in coefficients of correlation functions $B_{1/2}$ and ${C_i}$, which are related to linear terms in $S_2$ and $F^2$, we can put these to be zero when we evaluate the minimum of the potential,
$${
V(S_2=0, F^2=0)={1\over g_{11}} \big|\partial_1 W \big|^2.
}$$
To find the minimum it is useful to expand the prepotential ${\cal F}_0$ for $A_2$ geometry as
$${
{\cal F}_0=\sum_{b=0}^{\infty} {S_2}^b {\cal F}_{0,b}(S_1),
}$$
where we ignored $S_1$ independent part, which can be combined with the classical action of $SU(N_f)$ and construct the one-loop running coupling constant. In the matrix model computation, ${\cal F}_{0,b}$ are contributions of diagrams with $b$ boundaries, which are perturbatively calculable order by order. With this expansion, the superpotential and metric become
\begin{eqnarray}
W(S_2=0)&=&\alpha_1 S_1 +N_c {\partial {\cal F}_{0,0}(S_1) \over \partial S_1 }+N_f {\cal F}_{0,1}(S_1), \nonumber \\
g_{11}(S_2=0)&=&  {\rm Im} {\partial^2 {\cal F}_{0,0} \over \partial S_1 \partial S_1}. \nonumber
\end{eqnarray}
In our setup, the disk and annulus amplitudes are exactly known, \cite{Areview,AFOS,Fuji}
\begin{align}
2\pi i{\cal F}_{0,1} &= S_1 \bigg(\log {\Delta +\sqrt{\Delta^2- {4S_1 \over m_1 }} \over 2 \Lambda_0 } +{\Delta \over \Delta+\sqrt{\Delta^2-{4S_1\over m_1}} } -{1\over 2}\bigg) \nonumber \\
&\simeq S_1\log {\Delta \over \Lambda_0}-{S_1^2\over 2 m_1 \Delta^2}+\cdots, \nonumber \\
2\pi i{\cal F}_{0,2}&= {1\over 2} \log \bigg( \Delta +\sqrt{\Delta^2 -{4S_1 \over m_1} }   \bigg)     -{1\over 2}  \log \bigg( 2\sqrt{\Delta^2 -{4S_1\over m_1}}\bigg) \nonumber \\
&\simeq  {S_1\over 2 m_1 \Delta^2}+\cdots \nonumber,
\end{align}
where $\Delta=a_1-a_2$. With these expressions, we see that $W(S_2=0)$ reproduces the superpotential in \eqref{flavorSup} in the limit $m_2\to \infty$ and $S_2 \to 0$.  At the leading order, the minimum of the potential is given by
\begin{equation}
\langle {S}_1\rangle^{|N_c|}= ({m}_1 {\Lambda}_0)^{|N_c|} \left({\bar{{\Delta}} \over \bar{{\Lambda}}_0 }  \right)^{N_f}e^{2\pi i \bar{\alpha}_1}. \label{vevS}
\end{equation}
Note that $N_f>0>N_c$. Since there is an exponential suppression factor, vev of $S_1$ exists in physical region, which we regard as a dynamical scale of the theory on the anti-D5 branes.

Expanding the potential \eqref{Poten} around the minimum we can read off coefficients of linear terms in $S_2$ which yield the gaugino mass term for $SU(N_f)$ part,
\begin{align}
{2\pi i \over g_{YM}^2}m_{\lambda} &={2\pi i F_1 \over 16 \pi^2} \left[ -|N_c| {\partial^2 {\cal F}_{0,1} \over \partial S_1 \partial S_1}+2N_f {\partial {\cal F}_{0,2} \over \partial S_1}   \right] \, \bigg|_{\langle S_1 \rangle} -{|F_1|^2 \over 32 \pi^2 i{\Lambda}^4}{\partial^2 {\cal F}_{0,1} \over \partial S_1 \partial S_1}\, \bigg|_{\langle S_1 \rangle},
 \nonumber \\
&\simeq  {1 \over 16\pi^2} \left[{|N_c| +N_f \over m_1 \Delta^2} F_1+{|F_1|^2 \over 2i\, m_1 \Delta^2 {\Lambda}^4 } \right],
\end{align}
where we supplied a dimensionful parameter $\Lambda$. In the field theory limit, we take the string scale to be infinity, keeping the scale $\Lambda$ finite which should be identified with the scale of $S_1$ in \eqref{vevS} in our model. The $\alpha_1$, which is the size of ${\bf P}^1$, also has to scale appropriately \cite{fieldlimit}\footnote{In the geometric engineering one focuses on the leading effect of the small paramter $\Lambda /M_{st}$. Geometric quantities scale with the small paramter, for example the potential scales $V\sim (\Lambda/M_{st})^4$. Thus the original string scale in the potential cancels and it becomes field theory scale vacuum energy $V\sim {\cal O}(\Lambda^4)$. }. The vev of $F^1$ is cut-off independent and a finite quantity in the limit,
$${
F^1=- g_{11}^{-1} \partial_1 W \big|_0 \simeq \beta \Lambda^4,
}$$
where the $\beta$ is defined\footnote{Note that without loss of generality we can take the phase of $\Delta /\Lambda_0$ to be real. With this normalizaiton, we defined the $\beta$. } 
 by $2i\, {\rm Im} \bar{\alpha}_1 \sim \beta \log \Lambda_0$, which encodes geometric data of the ${\bf P}^1$. 
On the other hand, another correlation functions can be read off from the linear term in $F^2$.
\begin{eqnarray}
2\pi i C_i(0)&=&{-2\pi i \over 16\pi^2 }{\rm Re}\left[{\rm Im}\left({\partial {\cal F}_{0,1}\over \partial S_1} \right) {\Lambda}^{-4}\bar{F}^1+|N_c| {\partial {\cal F}_{0,1} \over \partial S_1}-2N_f{\cal F}_{0,2}  \right]\bigg|_{\langle S_1 \rangle} , \nonumber \\
&\simeq & {1\over 16 \pi^2} \left[  {\rm Im}\left({\langle S_1 \rangle \over m_1 \Delta^2} \right) {{\Lambda}^{-4}}{\rm Re} {F}^1 +( |N_c|+N_f){\rm Re}\left( {\langle S_1 \rangle \over m_1 \Delta^2} \right) \right].
\end{eqnarray}

Finally let us comment on the diagrammatical computation of the gaugino mass. Although our present geometric configuration does not include an unstable mode, we do not know explicitly the UV Lagrangian for the brane/anti-brane system. Thus it is not easy to compute the correlation functions studied above from Matrix model computations directly. However, the flop of the ${\bf S}^2$ wrapping the anti-brane is a smooth process because its physical volume can never be zero \cite{ABF,ASBVII}. The new geometry yields the brane/brane configuration. The world volume theory on the branes is quiver gauge theory with a superpotential,
$${
W_{SUSY}={m_1 \over 2} {\rm tr} (\Phi_1-a_1)^2 +{m_2 \over 2}{\rm tr} (\Phi_2-a_2)^2+Q_{12}\Phi_2 Q_{21}+Q_{21}\Phi_1 Q_{12}.
}$$
Using this explicit Lagrangian and technology developed in \cite{DVI,DVIII,DVIV,DVV}, we can compute the non-perturbative effect from perturbative Feynman diagram computations. In fact, explicit formulae for ${\cal F}_{0,0}$, ${\cal F}_{0,1}$ and ${\cal F}_{0,2}$ have been perturbatively computed by this method.

\section*{Acknowledgments}

We would like to thank H. Goh, M. Ibe, J. Marsano, N. Seiberg and T. Yanagida for discussion. C.P. and J.S. thank the hospitality of the Institute for the Physics and Mathematics of the Universe at the University of Tokyo.

This work is supported in part by DOE grant DE-FG03-92-ER40701. The work of H.O. is also supported in part by a Grant-in-Aid for Scientific Research (C) 20540256 from the Japan Society for the Promotion of Science, by the World Premier International Research Center Initiative of MEXT of Japan, and by the Kavli Foundation. C.P. and J.S. are also supported in part by Samsung Scholarship.

%%%%%%%%%%%%%%%%%%%%%%%%%%%%%%%%%%%%%%%%%%%%%%%%%%%%%%%%%

%%%%%%%%%%%%%%%%%%%%%%%%%%%%%%%%%%%%%%%%%%%%%%
%
% Bibliography
%
%%%%%%%%%%%%%%%%%%%%%%%%%%%%%%%%%%%%%%%%%%%%%%

\end{document}